\documentstyle[aps,prb,preprint]{revtex}
\newcommand{\beq}{\begin{equation}}
\newcommand{\eeq}{\end{equation}}
\newcommand{\la}[1]{\label{#1}}

\newcommand{\ep}{\varepsilon}
\newcommand{\He}{{\rm He}}
\newcommand{\myvec}[1]{\displaystyle {\bf #1}}
\begin{document}
\title{
  The $^4$He tetramer ground state in the Faddeev-Yakubovsky
  differential equations formalism}
\author{
I.N. Filikhin, S.L. Yakovlev
}
\address{
Dept. of Math. $\&$ Comput. Physics,
Institute for Physics \\ Sankt - Petersburg State University
198504 Ul'yanovskaya 1, Petrodvorets, St. Petersburg, Russia
}
\author{V.A. Roudnev}
\address{
Physics Department, University of South Africa, P.O. Box 392, Pretoria 0003
}
\author{B. Vlahovic}
\address
{
Physics Department,
North Carolina Central University \\ 1801 Fayetteville St. Durham
27707 NC, and\\
Jefferson Lab, 1200 Jefferson Avenue, Newport News, VA 23606, USA
}
\date{\today}
\maketitle
\begin{abstract}
The characteristics of the four $^4$He atom cluster are investigated using
the differential equations for Yakubovsky components.
Binding energy, mean-square radius and density function are calculated for
the ground state.
The spatial properties of the cluster and its subsystems are studied.
\end{abstract}
\bigskip

\section{INTRODUCTION}
In the present work clusters of three and four helium atoms are
investigated. In recent years such systems had attracted growing
attention due to development of experimental techniques allowing
to observe and to study two and three-particle Helium clusters
\cite{He21,He22} and as a result of theoretical developments which
made possible the construction of realistic pair potentials {\it ab
initio}\cite{LMcL,Aziz91,VanMourik,TTY,Aziz79,Aziz87}.

The last two decades have been significant progress in the
investigation of few-body systems. This progress rests 
on the development of new techniques for solving Faddeev and
Yakubovsky (FY) equations\cite{MF,MY,SSK,CCG} for wave function
components and growing power of computer facilities that allowed to
reach higher accuracy in the branch of Monte-Carlo techniques. Though equivalent to Schr\"odinger equation the system
of FY equations has significant advantages over the Schr\"odinger
one. These advantages have their origin in a proper choice of wave function
decomposition into the components. The equations are constructed
so that only one pair potential enters the equation for a
particular component provided that the interaction in the system
is given in terms of pairwise potentials. This nice feature
leads to very simple boundary conditions for the
components and simplifies their numerical approximation significantly.
Traditionally Yakubovsky equations are used in nuclear
physics\cite{KharLev,MerkYakGign,SSK,CCG,NNNN,YakTMF,O16,1}. In
spite of strong mathematical background and effective numerical
techniques developed for Faddeev and Yakubovsky equations only a
few papers devoted to molecular physics exploiting the technique
of Faddeev or Yakubovsky equations are known. Most of them are
devoted to a
system of Helium
trimer\cite{Duffy,Glockle,Carbonell,Moto,RudYak2}. No four-body
calculations of molecular systems based on Yakubovsky equations
are known to the authors up to now. The aim of the present
paper is to take the first step towards wider exploiting the rigorous
and effective technique of Yakubovsky equations in molecular and
chemical physics.
%
%

The numerical scheme exploited in this paper to solve differential Yakubovsky
equations is the cluster reduction method (CRM) \cite{1} which opens
a way to construct small subspaces containing solutions of the
equations. This method developed and repeatedly applied to the
problems of nuclear physics has been employed in the present paper
to perform a calculation of bound state characteristics of a system of four
helium atoms. The method allows to obtain not only the estimations for the binding energy
of a four particle system but also the wave function of the system
using computer resources economically. Another important advantage of
CRM is applicability of the method to model-free calculations of
multichannel reactions in a system of four particles\cite{NNNN,O16,1}.
Being applied to systems of four helium atoms it can be used to
calculate the reaction rates for $\He_2 +\He_2\rightarrow \He_3+{\rm He}$
and $\He_2 +\He_2\rightarrow \He_3^{*}+{\rm He}$ processes. Calculating
$s$-wave model of $\He_4$ tetramer in this paper we make a step
towards investigation of reactions in  systems of four atoms on the
base of strongly grounded mathematically correct methods.

This paper contains three additional sections and a conclusion.
In section 2 the model is described and the main equations are given.
In the section 3 the method of solution is presented. Section 4
contains the results of numerical calculations.

\section{FORMALISM}

The calculations presented here were performed in the framework of
differential Yakubovsky equations (DYE) for four particles in
configuration space. The formalism of differential Yakubovsky
equations was developed by S.P.Merkuriev and S.L.Yakovlev
\cite{MY}. Here we only give a brief description of these
equations touching upon the approximations that we use and
emphasizing the advantages of using the DYE instead of the
Shr\"odinger equation in investigations of bound states of four
particles. Detailed and sound description of the equations and
asymptotic boundary conditions can be found in the original paper
\cite{MerkYakGign} and in more detail in the book  by
Merkuriev and Faddeev\cite{MF}.

When considering a system of four particles it is
convenient to use Jacobi coordinates. For identical particles
there are two types of Jacobi coordinates, which correspond to different
partitions of the four particle system into subsystems.
The first type corresponds to the partitioning into a three-particle cluster
and one separate particle (3+1 type).
The second one corresponds to the partitioning into two-particle clusters
(2+2 type). The explicit expressions for the Jacobi coordinates through
the particle coordinate vectors ${\bf r}_k$, $k=1,2,3,4$ are given by the formulas
\begin{equation}
    \begin{array}{l}
        \myvec{x}_{2,1} ={\bf r}_2 - {\bf r}_1 \ ,\\
        \myvec{y}_{3,21} =({\bf r}_1+{\bf r}_2)/2 -{\bf r}_3 \ , \\
        \myvec{z}_{4,321} =({\bf r}_1+{\bf r}_2+{\bf r}_3)/3 -{\bf r}_4 \ , \\
    \end{array}
\label{Jackobi1}
\end{equation}
for the first type of coordinates and
\begin{equation}
    \begin{array}{l}
        \myvec{x}_{1,2} ={\bf r}_2 - {\bf r}_1 \ , \\
        \myvec{y}_{3,4} ={\bf r}_4 -{\bf r}_3 \ , \\
        \myvec{z}_{34,12} =({\bf r}_1+{\bf r}_2)/2-({\bf r}_3+{\bf r}_4)/2 \ ,
    \end{array}
\label{Jackobi2}
\end{equation}
for the second one. The Jacobi coordinates for partitions of the same type
but with different distribution of particles among the clusters can be
obtained by cyclic permutations of the subscripts enumerating the particles in
(\ref{Jackobi1}) and (\ref{Jackobi2}).

Suppose the Hamiltonian of a system of four particles has the form
\begin{equation}
  H=H_0+\sum\limits_{\alpha} V(\myvec{x}_{\alpha})\ ,
\end{equation}
where $H_0$ is the Hamiltonian of the system of free particles,
$\alpha$ stands for two-particle subsystems of the four-body system
and $V(\myvec{x}_{\alpha})$ is the potential of the interaction in the
pair with index $\alpha$. For the systems of identical particles the
wave function $\Psi$ can be expressed in terms of two Yakubovsky
components  $U^k$, $k$=1,2. $U^1$ corresponds to the partition 3+1
and  $U^2$ corresponds to the partition 2+2. The expression of the total
wave function of the system in terms of Yakubovsky components reads
\cite{MerkYakGign,1}
\beq
\begin{array}{r}
\Psi =
(I+P^{+}+P^+P^++ P^{-})(I+P^+_4+ P^-_4)U^1+ \\
+(I+P^{+}_1+ P^{-}_1)(I+P^+P^+)U^2 \ .
\end{array}
{\label{WF}}
\eeq
Here $P^+$ ($P^-$) are the operators of cyclic (anticyclic) permutations of
four particles, $P_i^\pm$  correspond to cyclic permutations of three
particles with fixed $i$-th particle. The Yakubovsky components
$U^1$ and $U^2$ satisfy the following set of the equations \cite{MerkYakGign}:
\begin{equation}
\begin{array}{rcl}
(H_0+V(\myvec{x})-E)\ U^1 & + & V(\myvec{x})(P_4^{+}+P_4^{-})\ U^1 = \\
          = & - & V(\myvec{x})\left((P_1^{+}+P^{+}) \ U^1+
                    (P_1^{+}+P_4^{+})\ U^2\right) \ ,\\
(H_0+V(\myvec{x})-E)\ U^2 & + & V(\myvec{x})(P^{+}P^{+})\ U^2 = \\
          = & - & V(\myvec{x})(P^{+}+P_1^{+})P^{+}\ U^1 \ .
\end{array}
\label{eqYack}
\end{equation}
Here we have omitted the subsystem index $\alpha$ in the notation of
coordinates $\myvec{x}$ since all the particles are identical.
The advantages of using the equations for Yakubovsky components
instead of solving directly the Shr\"odinger equation come from better
localization of the interaction in configuration space. In the special case
of identical particles the DYE can be written in term of the interaction
potential of only one pair. As a result the numerical approximation
of the Yakubovsky components is a much easier problem than the
approximation of the wave function. Detailed
discussion of Yakubovsky equations can be found in the monograph by
Merkuriev and Faddeev\cite{MF}.

We solve the equations for Yakubovsky components in the $s$-wave approximation
in which the angular momenta of the system of four atoms and all its subsystems
are set to zero. The $s$-wave equations for Yakubovsky components ${\cal U}^{k}$,
$k$=1,2  have the following form \cite{MF}
\beq
\begin{array}{lll}
(h^{1}_{0}\ +\ v(x) & - &\ep)\ {\cal U}^{1} (x,y,z) \ + \
  v(x)\displaystyle \int\limits_{-1}^{1}dv\ \frac{xy}{{x_1}{y_1}}
                                \  {\cal U}^{1}({x_1},{y_1},{z_1})= \\
                & = & -\displaystyle \frac{1}{2}v(x)
        \int\limits_{-1}^{1}du\ \int\limits_{-1}^{1}dv \
                    (\frac{xyz}{{x_2}{y_2}{z_2}}\ {\cal U}^{1}({x_2},{y_2},{z_2})\ + \\
 & & \ \ \ \ \ \ \ \ \ \ \ \ \ \ \ \ \ \ \ \ \ \ \ \
 +\ \displaystyle \frac{xyz}{{x_3}{y_3}{z_3}}\ {\cal U}^{2}({x_3},{y_3},{z_3}) ) \ , \\
(h^{2}_{0}\ +\ v(x) & - & \ep)\ {\cal U}^{2} (x,y,z)\ +\ v(x)\ {\cal U}^{2}(y,x,z)= \\
 & = & -\displaystyle \frac{1}{2}v(x)\int\limits_{-1}^{1} du\ \frac{xyz}{{x_4}{y_4}{z_4}}
                                                \ {\cal U}^{1}({x_4},{y_4},{z_4}) \ ,
\la{ya}
\end{array}
\eeq
\noindent
where $x=|{\bf x}|$, $y=|{\bf y}|$, $z=|{\bf z}|$,
$$
h^{1}_{0} = -( \partial^{2}_{x} + \frac{3}{4}\partial^{2}_{y} +
\frac{2}{3}\partial^{2}_{z} ),
$$
$$
h^{2}_{0} = -( \partial^{2}_{x} + \partial^{2}_{y} +
\frac{1}{2}\partial^{2}_{z} ),
$$
\noindent
$v(x)$ is s-wave component of the pair potential $V(x)$. The
coordinates $x_i,y_i,z_i$, $i=$1,2,3,4 in the kernels of the
equations (\ref{ya}) are defined by the following relations: $$
x_1=
\left(\frac{x^{2}}{4} + y^{2} + xyv\right)^{1/2}, \qquad y_1 =
\left(({\frac{3}{4}x})^{2} + \frac{y^{2}}{4} - \frac{3}{4}xyv\right)^{1/2},
$$
$$ {x_2} = {x_1},
\ \ {x_3} = {x_1}, \ \ {x_4} = y, $$ $$ {y_2} =
\left((\frac{y_1}{3})^{2}+z^{2}+\frac{2}{3}{y_1}zu\right)^{1/2}, \
\ {z_2} =
\left((\frac{8}{9}y_1)^{2}+\frac{z^2}{9}-\frac{16}{27}{y_1}zu
\right)^{1/2},
$$
$$
{y_3} = \left((\frac{2}{3}y_1)^{2}+z^{2}+\frac{4}{3}{y_1}zu\right)^{1/2}, \ \
{z_3} = \left((\frac{2}{3}y_1)^{2}+\frac{z^2}{4}-\frac{2}{3}
{y_1}zu\right)^{1/2},
$$
$$
{y_4} = \left((\frac{x}{2})^{2}+z^{2}-xzu\right)^{1/2}, \ \
{z_4} = \frac{2}{3} \left(x^{2}+{z^2}+2xzu\right)^{1/2}.
$$

\section{METHOD OF SOLUTION}

The differential equations for the Yakubovsky components
(\ref{ya}) are solved using the cluster reduction method (CRM).
This method has been developed and applied before
\cite{NNNN,O16,1} to calculate the characteristics of bound states
and low-energy scattering of systems of three and four particles.
The cluster reduction method reduces considerably the
computational difficulties when solving DYE numerically. In the
framework of the CRM Faddeev (Yakubovsky) components are
decomposed in terms of the eigenfunctions of the Hamiltonians of
two (three) particles subsystems. As a result of the projection
onto the elements of a biorthogonal basis we obtain the set of
equations corresponding to the relative motion of clusters. A
brief summary of the CRM from Yakovlev and Filikhin~\cite{1} is
given below. The Yakubovsky components ${\cal U}^{i}$, $i$=1,2 are
written in the following form
\beq
{\cal U}^{i}(x,y,z)= \sum\limits_{l=0}^{\infty }
\phi_{l}^{i}(x,y)F^{i}_{l}(z), \
\ \ i=1,2 .
\la{E5}
\eeq
In the Eq. (\ref{E5}) the basic functions
$\phi_{l}^{i}$ are the solutions of $s$-wave Faddeev
equations for subsystems of types $3+1$ $(i=1)$ and $2+2$
$(i=2)$:
\beq
\begin{array}{c}
 \{ -\partial_{x}^{2} -\frac{3}{4}\partial_{y}^{2}  +v(x)\}
          \phi^{1}_{l}(x,y) +
          v(x)\displaystyle\int\limits_{-1}^{1} dv
                  \frac{xy}{x_{1}y_{1}}\phi_{l}^{1}(x_{1},y_{1})= \\
      =\ep^{l}_{1}\phi_{l}^{1}(x,y)\ , \\
 \{ -\partial_{x}^{2}-\partial_{y}^{2}+v(x)\}
  \phi_{l}^{2}(x,y)
  +v(x)\phi_{l}^{2}(y,x) =\ep_{2}^{l}\phi_{l}^{2}(x,y) \ .
\end{array}
\la{fad}
\eeq

The set of functions $\{ \psi_{l}^{i}\}$ biorthogonal to the set
$\{ \phi_{l}^{i}\}$ consists of the eigenfunctions of the  equations
adjoint to the Eq. (\ref{fad})
\begin{equation}
\begin{array}{c}
\{ -\partial_{x}^{2} -\frac{3}{4}\partial_{y}^{2}  +v(x)\}
                                                                                                                \psi^{1}_{l}(x,y)
 + \int\limits_{-1}^{1} dv \frac{xy}{x_{1}y_{1}}v(x_{1})\psi_{l}^{1}(x_{1},y_{1})= \\
=\ep^{l}_{1}\psi_{l}^{1}(x,y) \ , \\
\{-\partial_{x}^{2}-\partial_{y}^{2}+v(x)\}
\psi_{l}^{2}(x,y)
+v(y)\psi_{l}^{2}(y,x) =\ep_{2}^{l}\psi_{l}^{2}(x,y).
\end{array}
\end{equation}
A biorthogonal basis is required because the Faddeev operator is
not self-adjoint \cite{YakTMF}. Substituting (\ref{E5}) into the
Eq. (\ref{ya}) and projecting onto conjugated basis $\{
\psi_{l}^{i}\} $ we obtain the set of integro-differential
equations for the functions $F_{l}^{i}(z)$, describing the
relative motion of clusters
\beq
\begin{array}{c}
\{ -\frac{2}{3}\partial_{z}^{2} +\ep_{1}^{l}-\varepsilon\} F_{l}^{1}(z) = \\
    = -\frac{1}{2} \sum\limits_{k=0}^\infty
          \left \langle \psi_{l}^{1}(x,y)|v(x)\int\limits_{-1}^{1}du
        \int\limits_{-1}^{1} dv
        \left \{ \frac{xyz}{x_{2}y_{2}z_{2}}
          \phi_{k}^{1}(x_{2},y_{2})F_{k}^{1}(z_{2}) + \right. \right. \\
      +\left. \left. \frac{xyz}{x_{3}y_{3}z_{3}}
        \phi_{k}^{2}(x_{3},y_{3})F_{k}^{2}(z_{3}) \right \} \right \rangle,
\\
\{ -\frac12\partial_{z}^{2} +\ep_{2}^{l}-\varepsilon\} F_{l}^{2}(z) = \\
 = -\sum\limits_{k=0}^\infty
      \left \langle \psi_{l}^{2}(x,y)|v(x)\int\limits_{-1}^{1}dv
      \frac{xyz}{x_{4}y_{4}z_{4}}\phi_{k}^{1}(x_{4},y_{4})F_{k}^{1}(z_{4})
      \right \rangle .
\end{array}
\la{E7}
\eeq
In these equations  $\langle .|. \rangle $ means the integration over the
variables $x$ and $y$.
The functions $F_{l}^{i}(z)$ must vanish when $z$ $\to$ $\infty$
$$
F_{l}^{i}(z) \sim 0, \qquad i=1,2, \qquad l=1,2,\dots,\infty.
$$
The number of equations in the set depends on the number of the
terms retained in the expansion of the Yakubovsky components, Eq. (\ref{E5}).
Due to completeness of the set of the basic functions only a finite number
$N$ of such terms
needs to be taken into account to support a stable numerical solution.

\section{RESULTS}

The solution of the Eq. (\ref{E7}) has been computed in the region
$\Omega$ of the configuration space defined by the
parameters  $R_{x}$, $R_{y}$, $R_{z}$:
$$
\Omega =\{x,y,z: x<R_x, y<R_y, z<R_z\}.
$$ The values of these parameters were chosen to be
$R_{x}$=$R_{y}$=$R_{z}$=50~\AA. All the calculations were
performed using model potentials HFDHE2\cite{Aziz79} and
HFD-B\cite{Aziz87}. From the one hand according to contemporary point
of view these potentials give lower
and upper limits for two-body binding energies correspondingly,
and from another hand some four-body results for these potential
models are known in literature\cite{NLim}.

The basic functions $\phi_{l}^{i}(x,y)$, $\psi_{l}^{i}(x,y)$,
$i=1,2$, $l=1,2,\dots,N$  were calculated using the CRM~\cite{1}.
To confirm the accuracy of the basic functions we were checking the
basis for orthogonality using the condition
$$ ( \langle \phi_{l}^{i} |
\psi_{m}^{i}\rangle-\delta_{lm}) < 10^{-3},\qquad i=1,2.
$$
The
function $\phi_{1}^{1}(x,y)$ for $l=1$ and $k=1$ in the Eq.
(\ref{E5}) is the $s$-wave Faddeev component of the ground state
wave function of the He$_3$ system (trimer). The binding energy of
the He$_3$ ground state when computed for the HFDHE2 and the HFD-B
potentials has the values -0.105 K and -0.118 K. Compared to the
values reported by Carbonell et al.\cite{Carbonell} and Kolganova
et al.\cite{Moto} our trimer is slightly overbound.

The He$_4$ (tetramer) binding energy, which has been computed
using the potentials HFDHE2 and HFD-B, is given in the Tab.~1. In
the same table we also quote the result reported by Nakaichi-Maeda
and Lim\cite{NLim}. These authors used the formalism of integral
AGS equations\cite{AGS}. In addition we include the results of the
calculations of the mean square radius ($<r^2>^{1/2}$) of the
system, the mean square distance between Helium atoms, and the
probability of forming the subsystem with cluster structure He$_3$
+ He. This last probability has been computed as
 \begin{equation}
 P_{3+1}=<\psi_1F_1|\Psi>,
 \end{equation}
where $\Psi$ is the total wave function of the system, $\psi_1$ is the
ground state wave function of He$_3$ and $F_1$ is the function describing
the motion of He$_3$ trimer relative to a single He atom. As can be seen
from the table, the contribution of the He$_3$ + He state to the total
wave function is considerable.

The fast convergence of the cluster decomposition, Eq. (\ref{E5})
indicates the existence of clusters in subsystems. Particularly,
in the 3+1 subsystem one needs to take into account two terms to
get a stable binding energy. The binding energy computed by
taking into account only the 3+1 component has the value
-0.25 K, which is in good agreement with the value -0.24 K
by Nakaichi-Maeda and Lim\cite{NLim}. To achieve a binding energy
calculation that is stable at the scale
$10^{-2}$~K one needs to take into account six terms in the
Eq. (\ref{E5}) which use components of both types (3+1) and (2+2).

The analysis of the results of our computation of the
characteristics of He$_3$ and He$_4$ systems enables us to draw an
analogy with the nuclear cluster systems 3$\alpha$ and 4$\alpha$.
Here the symbol $\alpha$ denotes a $^{4}$He nucleus. The bound
states of these systems correspond to the ground states of nuclei
$^{12}$C and $^{16}$O, respectively. It was known (for example
\cite{O16,Dub}) that in these cases the three-body systems 
have no well defined clusters in their subsystems. However, in the
four-body systems the cluster of 3+1 type is dominant i.e. in this
case it is possible to separate a closely bounded cluster of three
particles and fourth particle. The mean square radius of the
nuclear systems 3$\alpha$ ($<r^2>^{1/2}$=2.33 fm) and 4$\alpha$
($<r^2>^{1/2}$=2.54 fm)
\cite{O16} increases with the number of particles.
The situation is similar for the system under consideration. In
particular for the potential
HFDHE2 (HFDH-B) the mean square radius of trimer is 6.7~\AA (6.5~\AA)
and that of the tetramer is 7.4~\AA (6.9~\AA) for the same potentials.

To characterize the spatial distribution of the particles
constituting the tetramer we have computed its wave function
(\ref{WF}) for the HFDHE2 potential. The density function  $\rho(
r )$  is depicted in Fig.~1. normalized with the usual condition
$
\int\limits_0^{\infty}\rho(r)dr=1.
$

To study the spatial position of the Helium atoms in the tetramer
we have used the total wave function of the system. For comparison
analogous calculations were performed for the trimer. The He$_4$
(He$_3$) wave function depends on six (three) variables. These are
moduli of the Jacobi coordinates $x$, $y$, $z$ ($x$, $y$) and
cosines of the angles between vectors $\bf x$, $\bf y$, $\bf z$
($\bf x$ and $\bf y$) $u=\frac{({\bf x},{\bf y})}{xy}$,
$v=\frac{({\bf x},{\bf z})}{xz}$, $w=\frac{({\bf y},{\bf
z})}{yz}$. The most probable configurations of the relative
position of the particles forming the He$_4$ (He$_3$) system was
calculated as the coordinates of the maximum of the square of the
total wave function. For  He$_3$ system we found $x$=3.6~\AA,
$y$=3.1~\AA, $u$=0, and for He$_4$ system $x$=3.6~\AA,
$y$=3.1~\AA, $z$=2.9~\AA, $u$=0, $v$=0, $w$=0. These
configurations are shown in Fig.~2. For the ground state of the
trimer the center of the He atoms masses arrange themselves at the
vertices of the equilateral triangle with sides as large as
3.6~\AA (Fig.~2.~a). For the ground state of the tetramer the
three Helium atoms are located at the vertices of an equilateral
triangle with sides as large as 3.6~\AA, while the most probable
position of the fourth Helium atom is at a distance of 2.9~\AA ~in
the direction perpendicular to the plane of the three particle
system  (Fig.~2.~b) and through the center of the equilateral
triangle formed by them. One should not be surprised by the
predominance of the tetrahedron configuration if one takes into
consideration the identity of particles and the $s$-wave approach
that has been used for the description of the tetramer. One can
compare the location of density function maximums with the
positions of potential energy minimums. These positions differ
noticeably that demonstrate the essentially quantum nature of the
system. The minimums of potential forms a configuration of
equilateral tetrahedron with the side of 3.0~\AA whereas the
maximums of density function are located on the vertices of
tetrahedron with the side of 3.6~\AA.

\section{CONCLUSION}

By applying the method of cluster reduction we have solved
numerically  the s-wave differential equations  for the Yakubovsky
components for a system with four $^4$He atoms. Binding energy,
mean-square radius and density function are calculated for the
ground state. The results of the calculations are in good
agreement with those of Nakaichi-Maeda and Lim\cite{NLim}, which
were performed using the integral equations. The configurations
with He$_3$ cluster and separated helium atom dominates in the
He$_4$ cluster. This behavior is analogous to that of the  nuclear
4$\alpha$ particles system \cite{O16}. The most probable spatial
configuration of the four Helium atom system is the tetrahedron
with sides as large as 3.6~\AA.

\acknowledgments
The authors wish to thank V.M. Suslov for useful discussions, A. Soldi for
useful suggestions and the North Carolina Supercomputing Center for CPU time.
S.L.Y. and I.N.F. would like to thank the Russian Foundation for
Basic Researches (grant No. 98-02-18190) and the Russian Ministry of Education
(grant No. 97-0-14.3-23) for financial support. I.N.F. would like to thank
Jefferson laboratory, Duke University - TUNL, Old Dominion University, Hempton
University, and
North Carolina Central University, for financial support and hospitality.

\newpage
\begin{figure}[t]
\caption{
The comparison of the probability densities for three- and four
He atom system: solid curve corresponds to
the ground state of the He$_4$ system,
dashed curve corresponds to the ground state of the He$_3$ system
(HFDHE2 potential).}
\end{figure}
\begin{figure}[t]
\caption{
The most probable configurations of Helium atoms:
a) the ground state of the He$_3$ system,
b) the ground state of the He$_4$ system.
The figures show the numbers of atoms and distances between
their centers of mass (HFDHE2 potential).}
\end{figure}
\newpage
\begin{table}
\caption{$^{4}$He$_4$ tetramer
binding energy ($E_4$), mean-square radius ($<r^2>^{1/2}$),
mean square distance between Helium atoms
($<x^2>^{1/2}$) and the contribution of cluster
subsystems of the He$_3$ + He form (P$_{3+1}$).}
\begin{tabular}{cccc}
  Potential          & \multicolumn{2}{c}{ HFDHE2} & HFD-B \\
\tableline
                                & present work & S. Nakaichi-Maeda and T.K. Lim\cite{NLim} & present work \\
\tableline
    $E_4$, K          & -0.39              &    -0.394 & -0.41            \\

$<r^2>^{1/2}$,~\AA &  7.4            &          --  & 6.9                \\

$<x^2>^{1/2}$,~\AA & 11.1            & --            & 10.3              \\

P$_{3+1}$              &  0.75              & --            & 0.81              \\
\end{tabular}
\end{table}
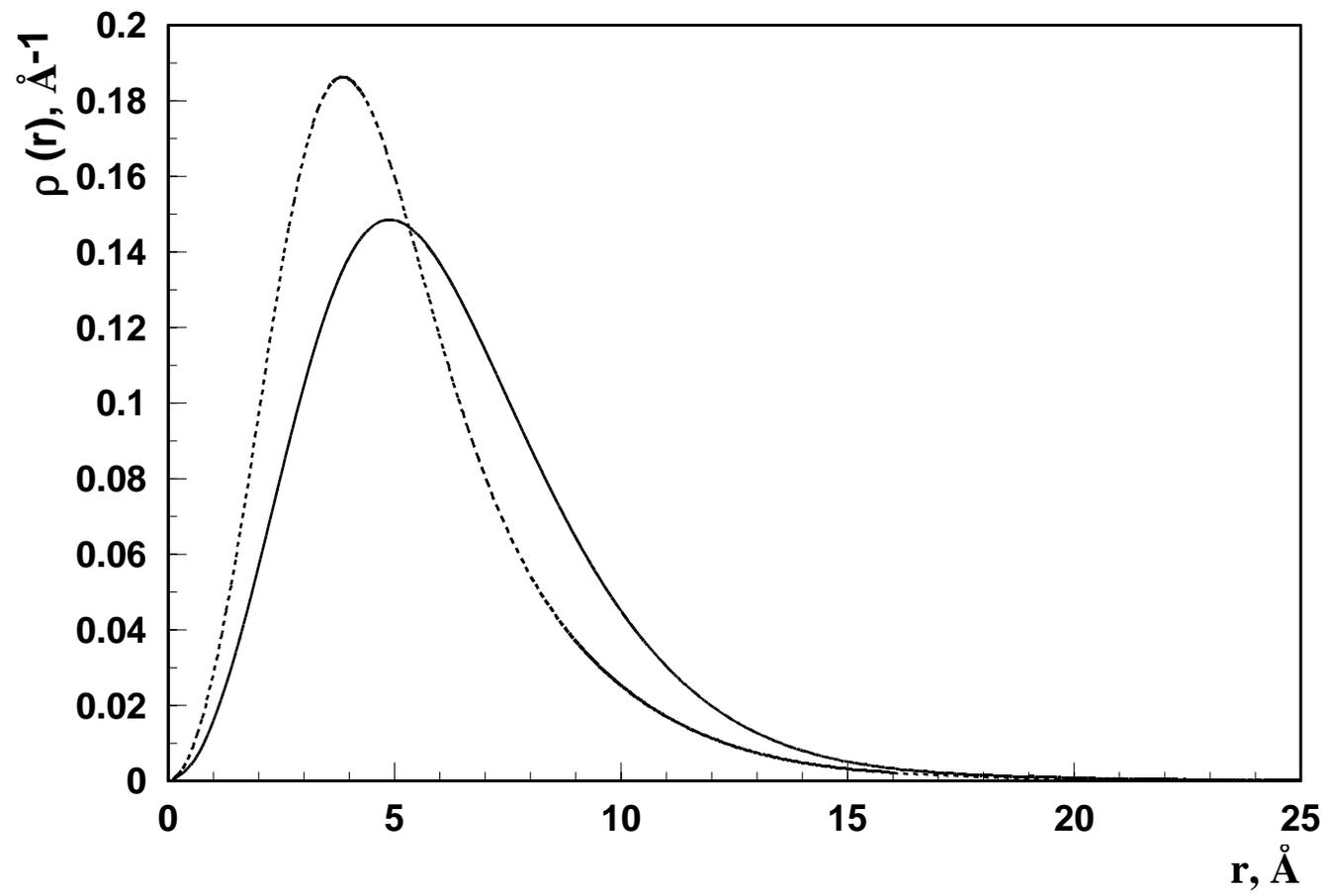
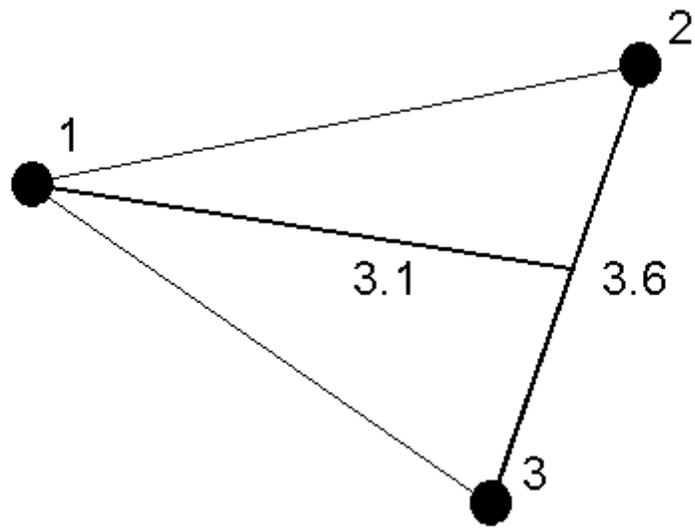
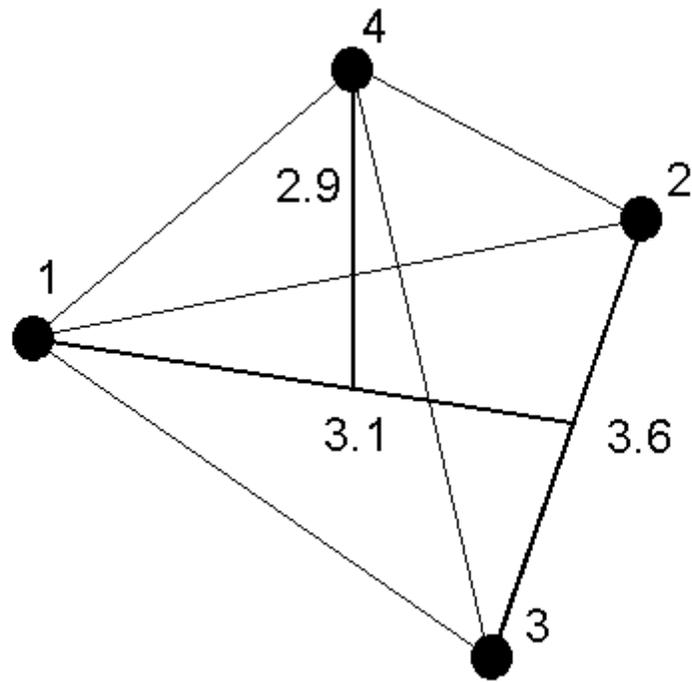
\end{document}